\documentclass{ws-ijmpcs}

\usepackage{epstopdf}

\begin{document}

\markboth{A.G. Pili, N. Bucciantini, L. Del Zanna}
{General Relativistic Equilibrium Models of Magnetized Neutron Stars}

%
\catchline{}{}{}{}{}
%

\title{GENERAL RELATIVISTIC EQUILIBRIUM MODELS OF MAGNETIZED NEUTRON STARS}

\author{A.~G. PILI}

\address{
Dipartimento di Fisica e Astronomia, Universit\`a di Firenze, \\
Via G. Sansone 1, 50019 Sesto F.~no  (Firenze), Italy\\
pili@arcetri.astro.it}

\author{N. BUCCIANTINI}

\address{INAF - Osservatorio Astrofisico di Arcetri\\
 L.go E. Fermi 5, 50125 Firenze, Italy\\
niccolo@arcetri.astro.it}

\author{L. DEL ZANNA}

\address{Dipartimento di Fisica e Astronomia, Universit\`a  di
  Firenze\\
 Via G. Sansone 1, 50019 Sesto F.~no  (Firenze), Italy}

\maketitle

\begin{history}
\received{Day Month Year}
\revised{Day Month Year}
\end{history}

\begin{abstract}
Magnetic fields play a crucial role in many astrophysical scenarios and,
in particular, are of paramount importance in the emission mechanism and
evolution of Neutron Stars (NSs). To understand the role of the magnetic
field in compact objects it is important to obtain, as a first step,
accurate equilibrium models for magnetized NSs. Using the conformally flat 
approximation we solve the Einstein's equations together with the GRMHD
equations in the case of a static axisymmetryc NS taking into account
different types of magnetic configuration. This allows us to
investigate the effect of the magnetic field on global properties
of NSs such as their deformation.

\end{abstract}

\ccode{PACS numbers: 97.10.Ld,  97.60.Jd, 04.25.D-}

\section{Introduction}	
Neutron Stars (NSs) are the most compact object of the universe that are
endowed with an internal structure, they can rotate very fast and harbour
very strong magnetic fields, up to $10^{16}$~G for magnetar.
Such type of magnetic fields are responsible for the phenomenology
of many NSs like  
Anomalous X-Ray Pulsars and Soft Gamma Repeater and they are invoked in
the {\it millisecond magnetar model} for
Long and Short Gamma Ray Bursts\cite{Bucciantini_Quataert+09a,Bucciantini_Metzger+12a}.
Finally strong magnetic field could deform the NS structure leading, together with
 rotation, to  gravitational wave emission. 

Here we present our work on equilibrium model for magnetized NSs.
Until now only simple configurations of
purely poloidal\cite{Bocquet_Bonazzola+95a} or purely toroidal magnetic 
fields\cite{Kiuchi_Yoshida08a,Frieben_Rezzolla12a}, have been
investigated in full GR. 
However more realistic magnetic configuration, 
the so called Twisted-Torus (TT),
require the presence of both poloidal and toroidal magnetic field. Lately TT models have 
been obtained either in the Newtonian regime \cite{Lander_Jones09a} or
using a perturbative
approach\cite{Ciolfi_Ferrari+09a,Ciolfi_Ferrari+10a,Ciolfi_Rezzolla13a}. Here
we present a  study (including TT cases) done in full GR and for
strong deformations.

\section{Assumptions for the equilibrium model}
In this work we consider only static stars (magnetar 
are typically slow rotators). We assume axisymmetry, 
ideal MHD, and a simple polytropic equation of state.
The metric is assumed to be {\it conformally flat} (CFC), such that in
spherical coordinates:
\begin{equation}
ds^2=-\alpha^2 dt^2+ \psi^4 (dr^2 + r^2d\theta^2 + r^2\sin{\theta}^2d\phi^2),
\end{equation}
where $\alpha$ is the lapse function, and $\psi$ is the conformal factor.
This approach allows us to simplify the Einstein's
equations, cast them in a numerically stable form and, as a 
consequence, handle stronger fields and deformations that in a 
perturbative approach can not be obtained, without compromising the 
accuracy of our results as we will discuss.
With this approximation the Eistein's equations reduce 
to the following Poisson-like equations:
\begin{equation}
\Delta \psi =-[2\pi\psi^6(e +B^2/2)]\psi^{-1},
\label{eq:1}
\end{equation}
\begin{equation}
\Delta(\alpha\psi)=[2\pi\psi^6(e +B^2/2) +2\pi\psi^6(6p+B^2)\psi^{-2}](\alpha\psi),
\label{eq:2}
\end{equation}
where $\Delta$ is the standard Laplacian operator in spherical coordinate while $e$, 
$p$ and $B^i$ are respectively the energy density, the pressure and 
the magnetic field as measured in the lab frame. 

The
equations of the static MHD give the Euler equation which, in the 
hypotheses of a polytropic EOS, can be 
integrated to obtain the equilibrium condition:
\begin{equation}
\ln \left(h/h_c\right)+\ln \left(\alpha/\alpha_c\right)-\mathcal{M}=0,
\end{equation}
where $h=(e+p)/\rho$ is the specific enthalpy, $\rho$  the rest mass density, $c$ 
denotes central values, and $\mathcal{M}$ is 
related to the Lorentz force by $L_i=\rho h\partial_i\mathcal{M}$. 
In the case of a purely poloidal magnetic field the integrabilitity of Euler equation
requires $\mathcal{M}$ to be a function of the fluid quantities $\rho$ and
$h$ and the metric functions $\psi$ and $\alpha$. When also a poloidal magnetic field is present
the solenoidality condition for the magnetic field together with  axisymmetry allows to rewrite
the poloidal components of the magnetic field in terms of derivative of only the $\phi$-component of
the vector potential $A_\phi$ while the toroidal component $B_\phi$ is related to $A_\phi$ by means of
a scalar function $\mathcal{I}$. Finally the metric and fluid quantities are related to the magnetic 
flux function $A_\phi$ by the Grad-Shafranov equation:
\begin{equation}
\tilde{\Delta}_3 \frac{A_\phi}{r\sin \theta} + 
\frac{\partial A_\phi \partial \ln(\alpha\psi^{-2})}{r\sin\theta}
+\psi^8 r \sin \theta \left( \rho h \frac{d \mathcal{M}}{d A_\phi}+ \frac{I}{\alpha^2\psi^4 r^2 \sin^2\theta}  \frac{d \mathcal{I}}{d A_\phi} \right),
\label{eq:3}
\end{equation}
where $\tilde{\Delta}_3$ is the $\phi$ component of the vector 
Laplacian in spherical coordinates. 

\section{Numerical Method}
The 
complete algorithm of the metric solver is fully presented in  
Ref.~\refcite{Bucciantini_Del-Zanna11a} (see also Ref.~\refcite{Del-Zanna:2007}).
The metric Eqs.~(\ref{eq:1}) (\ref{eq:2}) and 
the Grad-Shafranov Eq.~(\ref{eq:3}) are solved iteratively, because of 
non-linearity, exploiting a semi-spectral method: we decouple
the radial and angular dependency of the unknown function
$\alpha$, $\psi$ and $A_\phi$ using scalar and 
vector spherical harmonics.
Equations are reduced to a set
of ordinary differential equations for each harmonic that is solved with a
direct inversion of a tridiagonal matrix.
In all our models we have used 20 spherical harmonics for the elliptic
solvers and discretized the
domain $r=[0,30]\, \mbox{km}$, $\theta=[0,\pi]$, with 
250 points in the radial direction and 100 point in the angular one. 
At this resolution the accuracy of our results are 
of the order of $10^{-3}$.

\section{Results}

\begin{figure}[pb]

\begin{minipage}{0.99\textwidth}
\centerline{\psfig{file=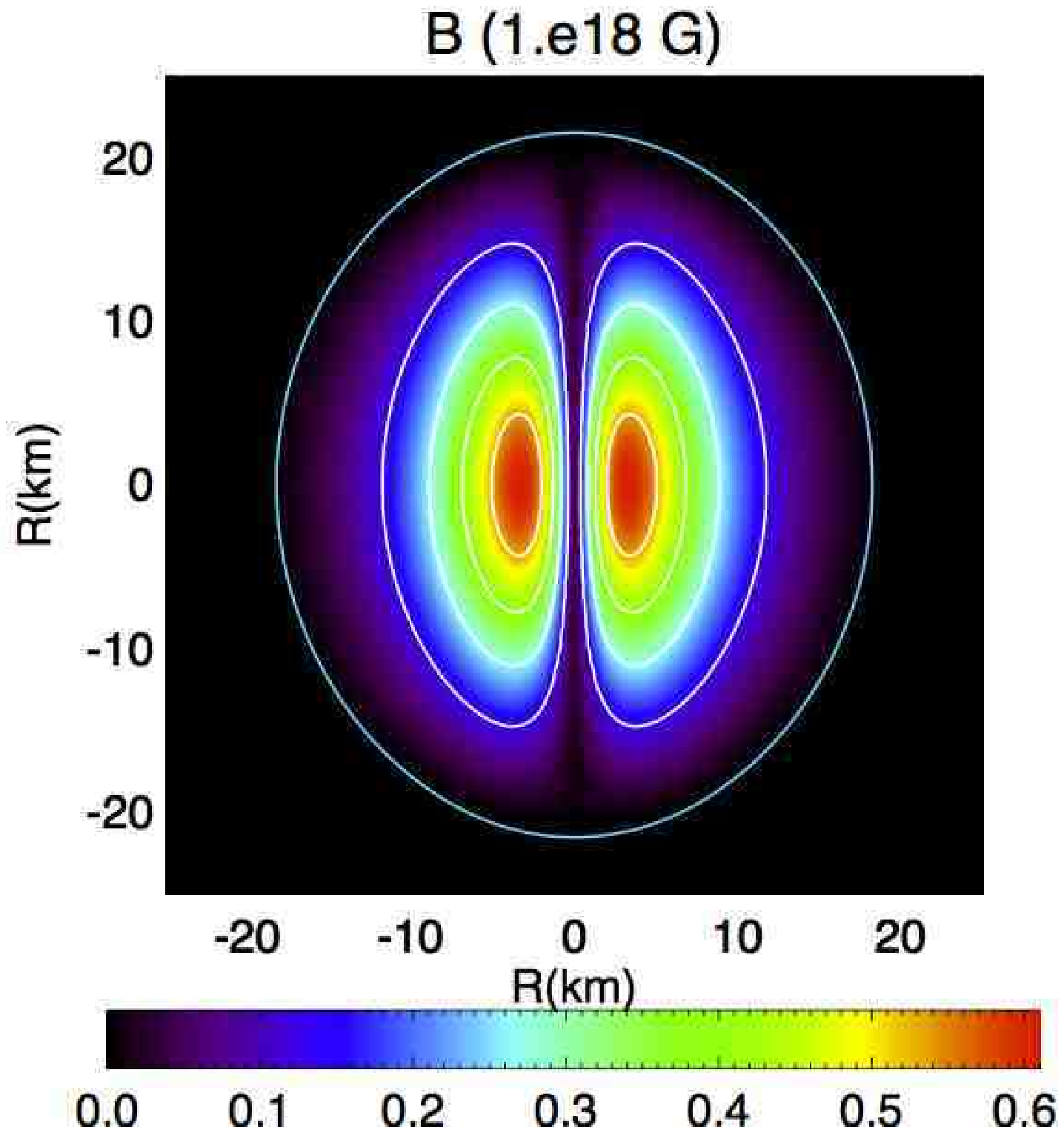,width=4.1cm, height=4cm}
\psfig{file=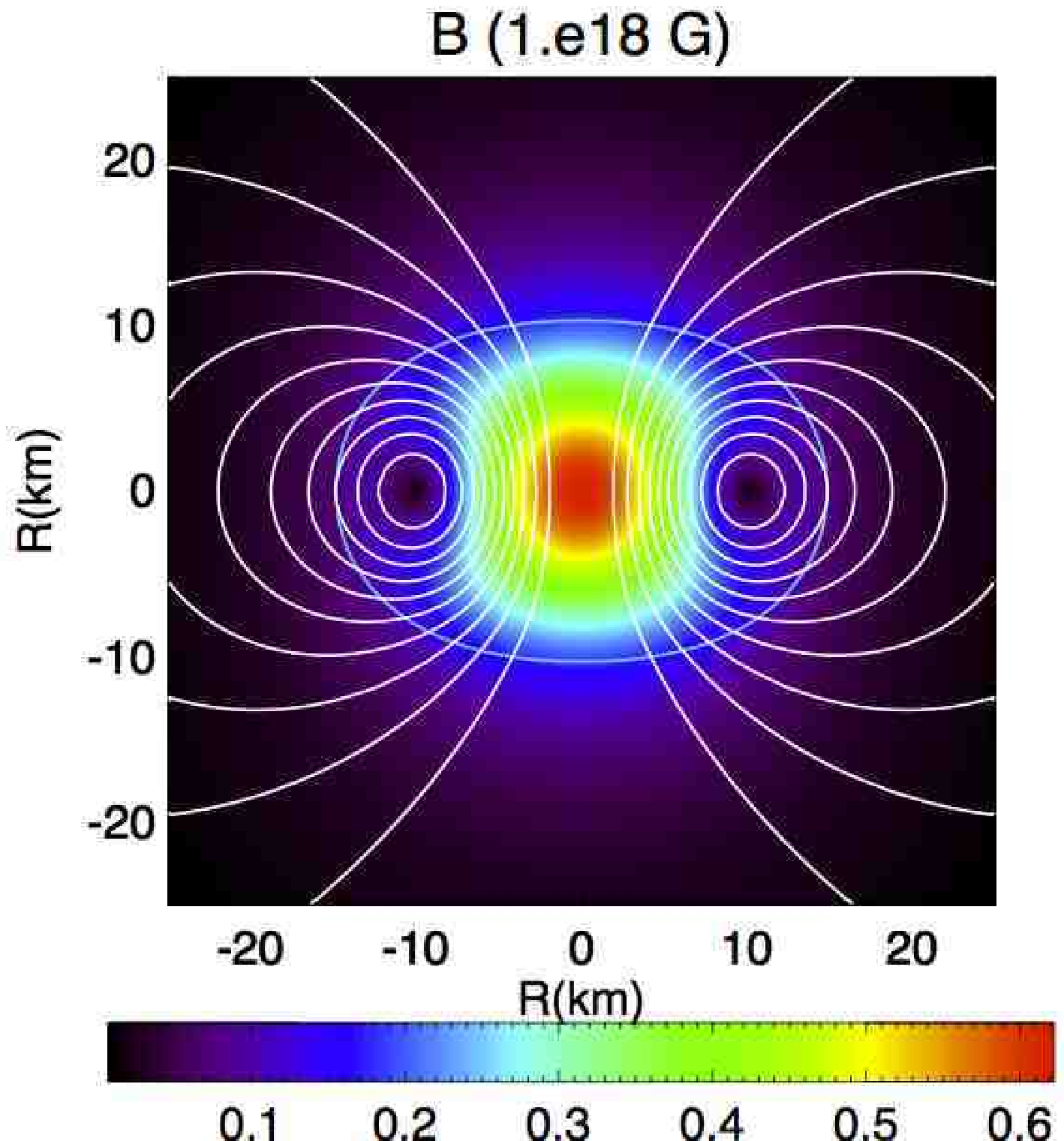,width=4.1cm,height=4cm}
\psfig{file=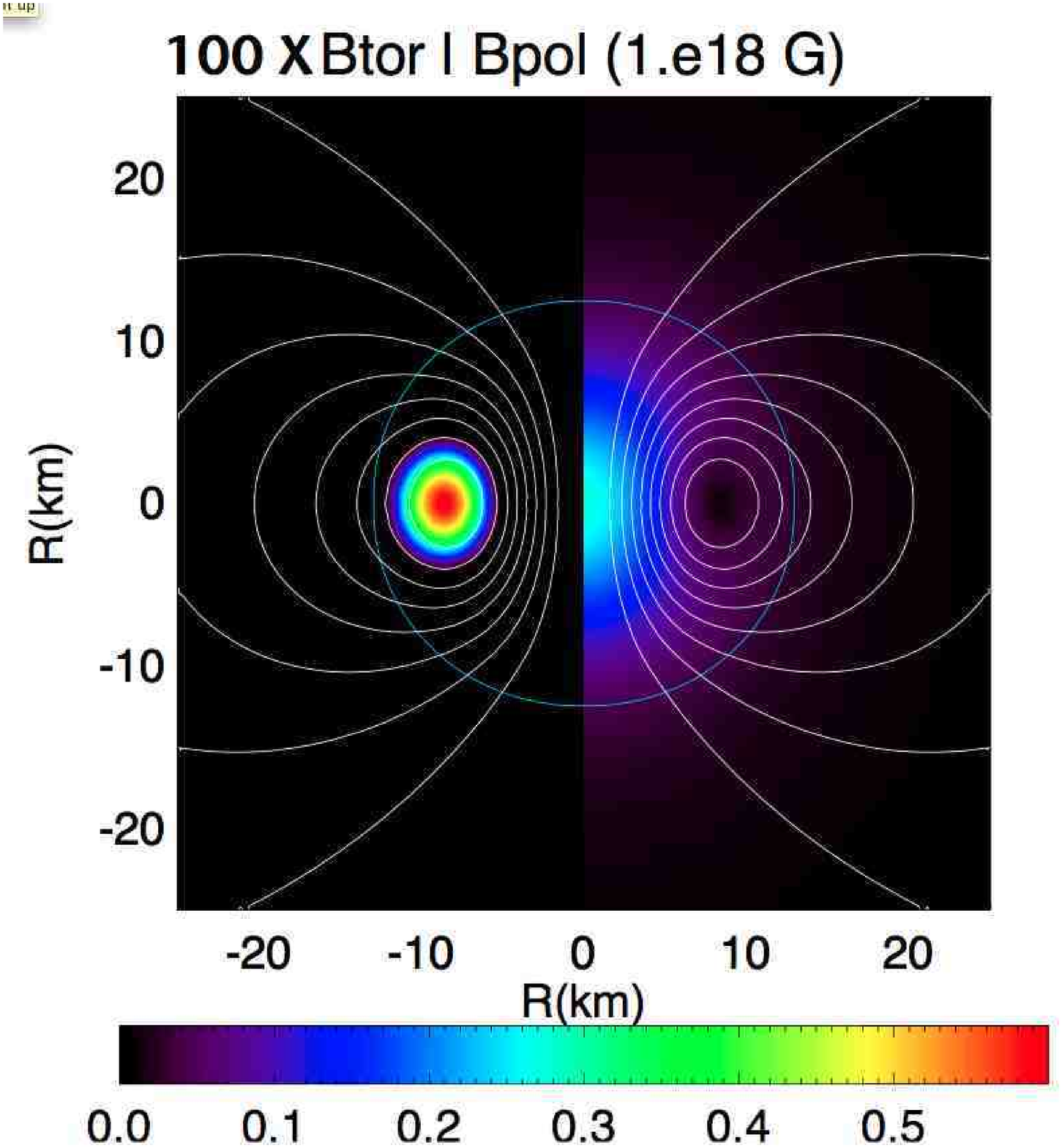,width=4.1cm,height=4cm}}\vspace*{18pt}
\end{minipage}\\\vspace*{18pt}
\begin{minipage}{0.95\textwidth}
\centerline{\psfig{file=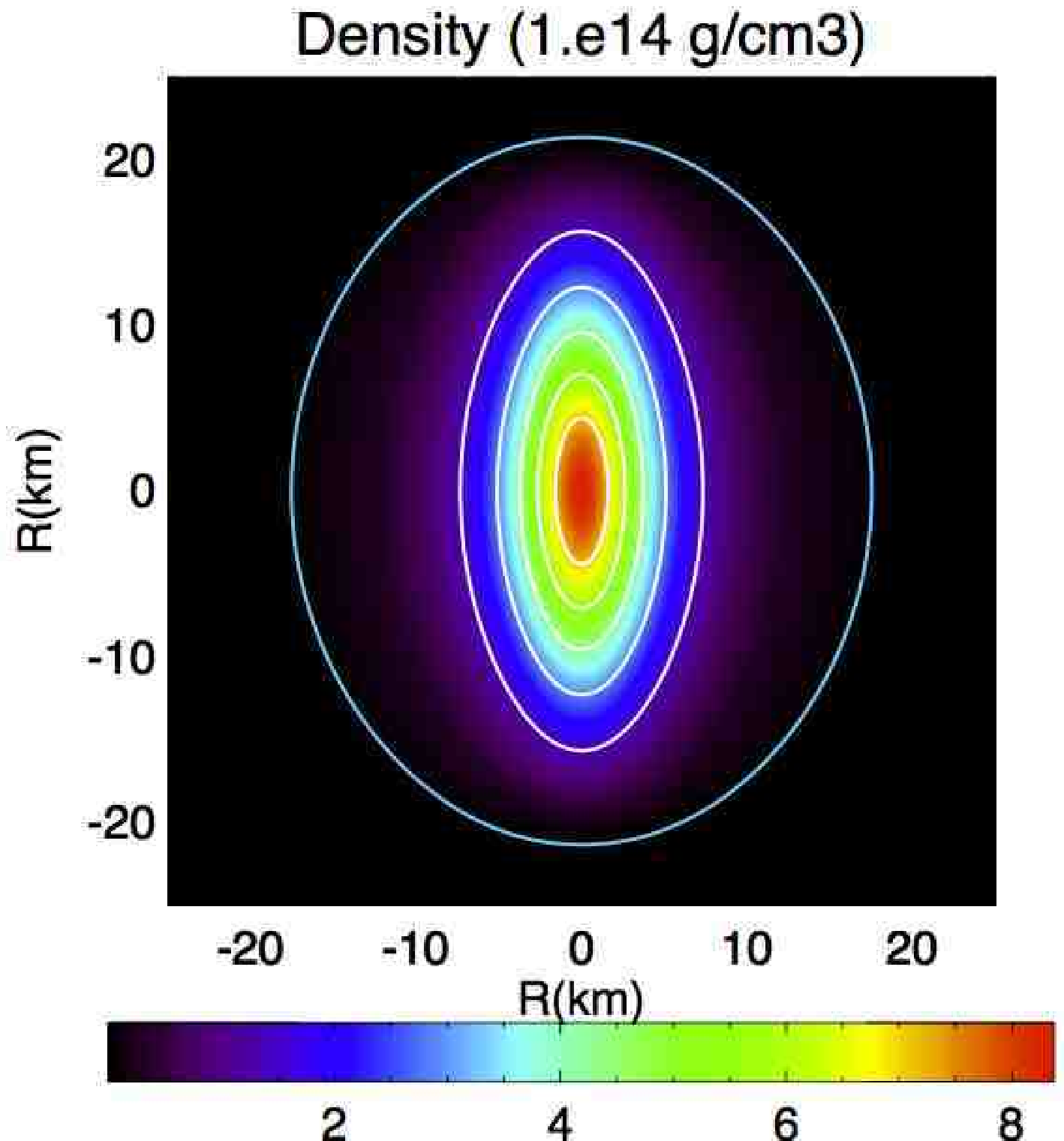,width=4.1cm}
\psfig{file=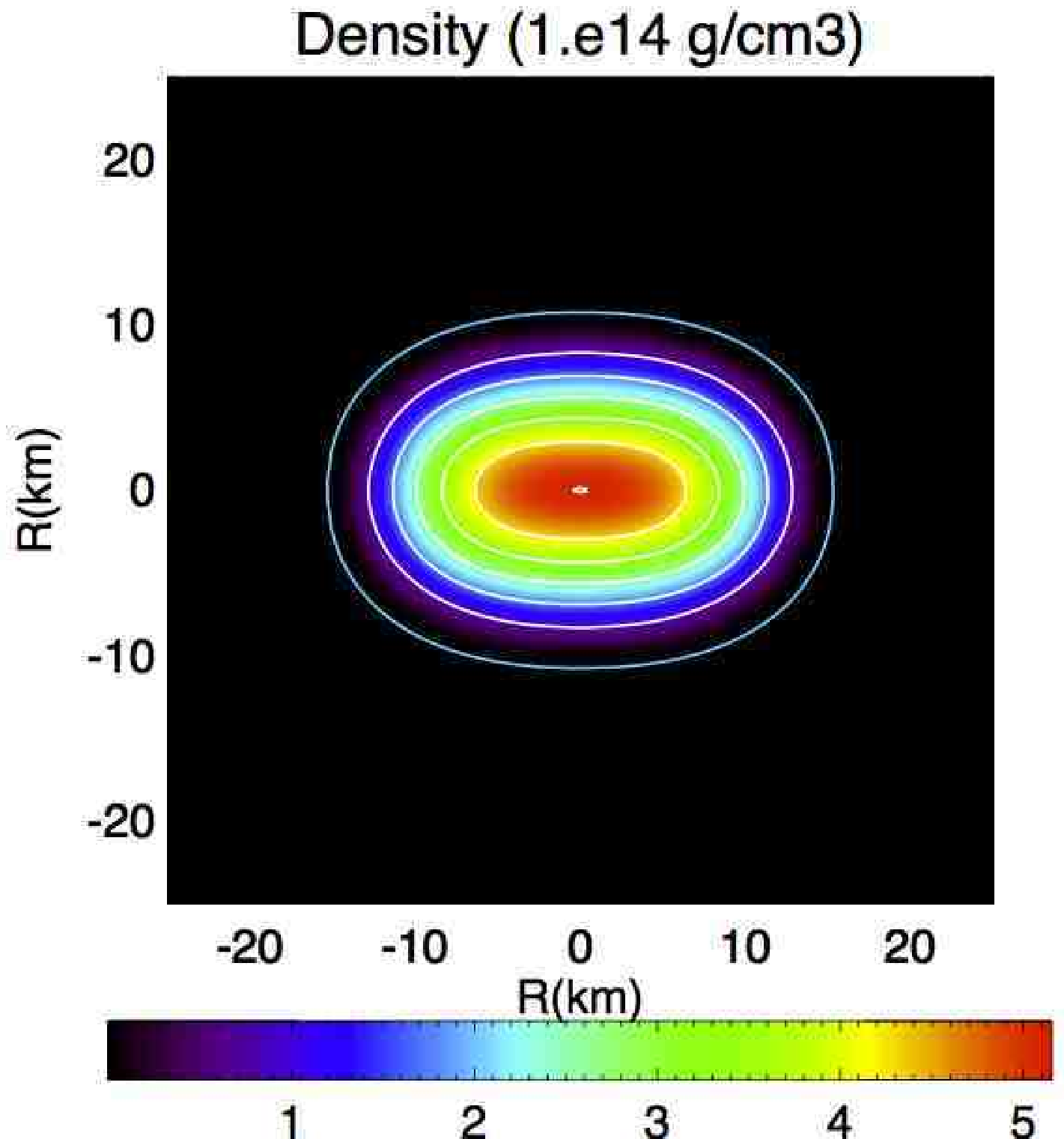,width=4.1cm}
\psfig{file=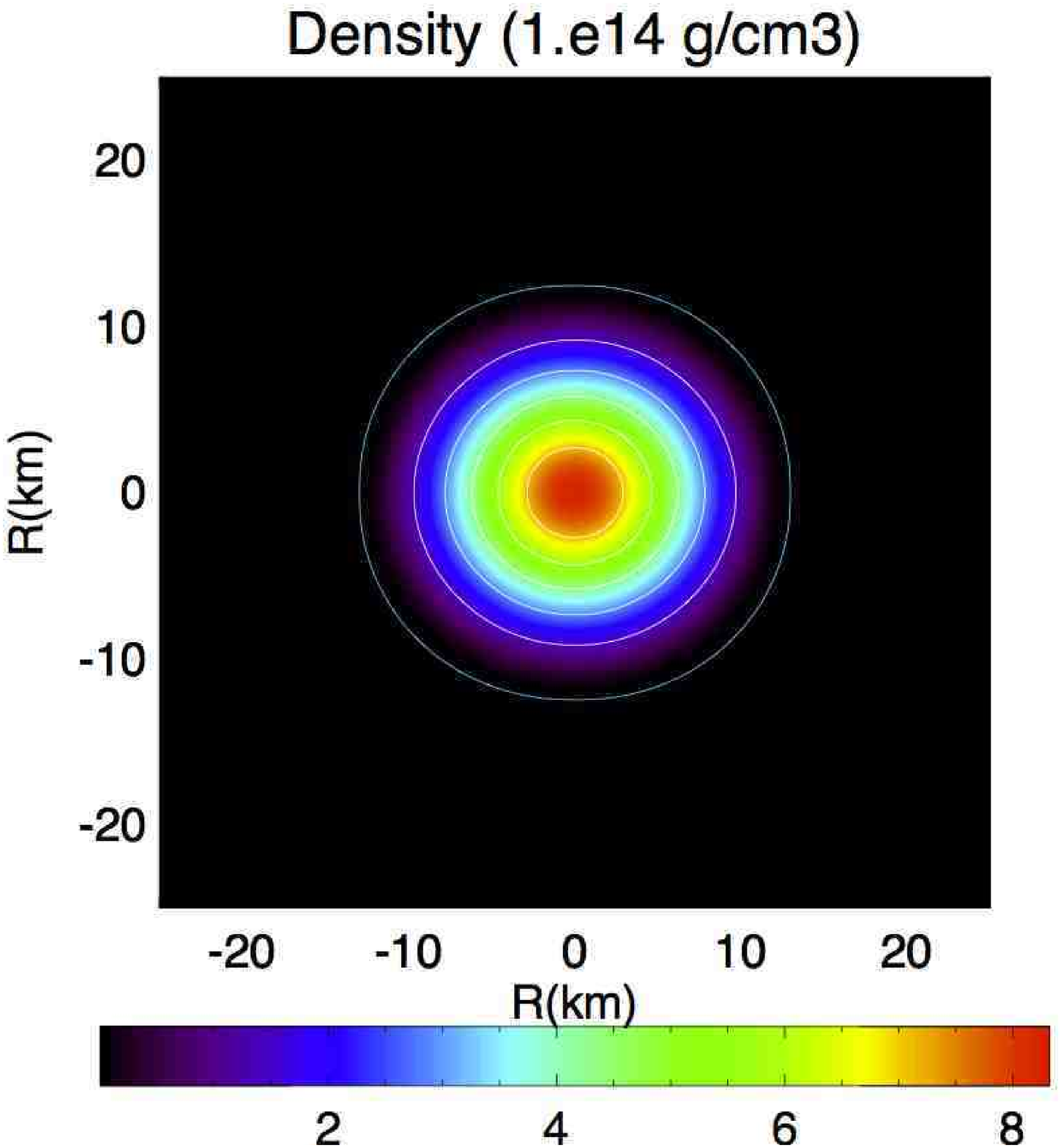,width=4.0cm}}
\end{minipage}
\caption{Magnetic field strength (top row), and density distribution
  (bottom row) for: a purely toroidal (left), a purely poloidal
  (central) and TT (right) magnetic configuration with $M_0=1.680 \,
  M_{\odot}$. Contours in the poloidal and TT case (upper panels) represents
  magnetic surfaces.  The blue line is the stellar surface. \label{f1}}
\end{figure}

With appropriate
choices of $\mathcal{M}$ and $\mathcal{I}$ it is possible to obtain either purely toroidal, 
purely poloidal and mixed field configurations.
A comparison between these magnetic configuration are 
shown in Fig.~\ref{f1}. 
In our models the toroidal magnetic field is always 
buried under the surface of the star while the poloidal magnetic 
field extends smoothly outside the star. In TT configurations 
the toroidal component does not pervade all the interior of the star 
but remains confined in a small ring-like region where the poloidal component vanishes.

A purely toroidal magnetic field induces a prolate deformation
while a purely poloidal one causes an oblate 
deformation of the star. In the TT configuration the poloidal field,
 usuaklly dominant over the toroidal
one, leads to oblate configurations. We found that the 
deformation is dominated by the field in the central region 
and is less sensible to peripheral magnetic fields and currents.
We built equilibrium sequences and explored
the parameter space of the different magnetic configurations.
The main results obtained are: 
\begin{itemlist}
\item the characteristic deformation induced by purely toroidal field
      is prolate: the magnetic stresses act concentrating the  
      central layers of the star around the 
      magnetic axis and causing an expansion of 
      the outer layers; 
\item a purely poloidal field leads to an oblate deformation, concentrating the core region in a disk-like	
      region orthogonal to the axis. In the most extreme
      cases we can also obtained torus-like density distributions;
\item magnetic fields and current concentrated in the outer 
      region have only marginal effects on the stellar structure; 
\item CFC results are in agreement with those obtained in the correct regime  
      \cite{Kiuchi_Yoshida08a,Frieben_Rezzolla12a,Bocquet_Bonazzola+95a};
\item in the TT configuration, that we have obtained for the first
      time in the fully non-linear regime, the toroidal 
      magnetic field is sub-dominant, and the deformation is due to the poloidal one that 
      acts deeper inside the star;
\item more compact configurations, with higher central rest mass 
	  density, can sustain higher magnetic field but exhibit minor 
	  deformations.

\end{itemlist}


\end{document}